\documentclass[doublecol]{epl2}

\title{Path diversity improves the identification of influential spreaders}
\shorttitle{} 

\author{Duan-Bing Chen\inst{1,2} \and Rui Xiao\inst{2} \and An Zeng\inst{2}\footnote{E-mail: an.zeng@unifr.ch} \and Yi-Cheng Zhang\inst{1,2}}
\shortauthor{D.-B. Chen \etal}

\institute{
  \inst{1} Web Sciences Center, School of Computer Science \& Engineering,  University of Electronic Science and Technology of China -  Xiyuan Avenue 2006, Chengdu 611731, People's Republic of China\\
  \inst{2} Department of Physics, University of Fribourg - Chemin du Mus\'{e}e 3, CH-1700 Fribourg, Switzerland
}

\pacs{89.75.-k}{Complex system}
\pacs{89.20.Ff}{Computer science and technology}
\pacs{89.65.Ef}{Social organization; anthropology}

\abstract{Identifying influential spreaders in complex networks is a crucial problem which relates to wide applications. Many methods based on the global information such as $k$-shell and PageRank have been applied to rank spreaders. However, most of related previous works overwhelmingly focus on the number of paths for propagation, while whether the paths are diverse enough is usually overlooked. Generally, the spreading ability of a node might not be strong if its propagation depends on one or two paths while the other paths are dead ends. In this Letter, we introduced the concept of \emph{path diversity} and find that it can largely improve the ranking accuracy. We further propose a local method combining the information of path number and path diversity to identify influential nodes in complex networks. This method is shown to outperform many well-known methods in both undirected and directed networks. Moreover, the efficiency of our method makes it possible to be applied to very large systems.}

\begin{document}

\maketitle
\section{Introduction}
How to identify influential nodes in complex networks is a crucial issue since it is highly related to the information spreading and epidemic controlling \cite{Kitsak-ranking-2010,Sinan2012,Bond2012,Frank2012}. So far, a number of centrality indices have been proposed to address this problem such as degree, betweenness \cite{Brands2001-JMC,Freeman1977}, closeness \cite{Noh-2004PRE,Opsahl-2010SN} and eigenvector centralities \cite{Katz1953}. Among these indices, degree centrality is a very straightforward and efficient one. However, the performance of degree centrality is not satisfying enough. Recently, Kitsak \emph{et al.} \cite{Kitsak-ranking-2010} claimed that the location of a node in the network actually plays a more important role  than the degree of it. They accordingly proposed a coarse-grained method by using $k$-shell decomposition to quantify a node's influence based on the assumption that nodes in the same shell have similar influence and nodes in higher shells are likely to infect more nodes. After this, some methods are proposed to further improve the ranking performance of this network decomposition process~\cite{PLA3771031,NJP14083030}. In directed networks, the ranking methods are mainly based on the iterative process. The representative methods include the well-known HITs \cite{Kleinberg-HITs-1999} and PageRank \cite{Page-PR-1999}, as well as some recently proposed algorithms like LeaderRank \cite{Lv-leader-2011} and TwitterRank \cite{TwitterRank2010}. It has been demonstrated that these methods outperform out-degree centrality in terms of ranking effectiveness.

With big data era coming, the design of ranking algorithms on very large-scale social networks is becoming a big challenge nowadays \cite{saito2012efficient}. The online social systems can have millions of users or even more. The spreader ranking algorithms will be very time-consuming if they are based on global information of the network. Therefore, the spreader ranking algorithm should be not only effective but also efficient. To solve this problem, it is better to design the ranking algorithm based on local information of the network. For example, a semi-local index by considering the second nearest neighbors is devised~\cite{Chen-ranking-2011}. This index is shown to well identify influential nodes and obtain a good trade off on effectiveness and efficiency comparing with global indices.

Moreover, most of previous ranking methods are designed based on the number of paths for propagation. Actually, the diversity of paths for spreading is also very important. The spreading ability of a node will be significantly lowered if many of its propagation paths overlapped. In this case, if the virus/information fails to go through the overlapped path, the following spreading of many paths will be terminated. However, this factor hasn't been taken into account in designing the spreading ranking algorithm so far, to the best of our knowledge.

Accordingly, we introduced in this Letter the concept of \emph{path diversity} \cite{PathDiver2003} which is mathematically characterized by the information entropy \cite{Shannon1948,VanSiclen1997PRE}. After applying it to a very simple spreader ranking method, we find that the performance of this ranking method can be further enhanced. Combining the information of path number and diversity, we further propose a local but effective ranking method (called KED method) to identify influential nodes in large scale social networks. We make use of the SIR spreading model \cite{Anderson-BOOK-1992} with tunable infectivity \cite{Zhou-PRE-2006,Yang-PLA-2007,Gomez-EPL-2010} to test the effectiveness of our method on four real social networks, including two undirected networks, Youtube and Orkut \cite{youtube-2012}, and two directed networks , EmailEU \cite{email-2007} and Digg \cite{digg-2010}. Experimental results show that KED performs much better than the simplest degree centrality, PageRank and LeaderRank, and in most cases better than $k$-shell, i.e. KED can more accurately rank the nodes on their correct places according to their real spreading ability than other ranking algorithms. In addition, the top-$L$ influential nodes identified by KED lead to much wider spreading than those by degree centrality, PageRank, LeaderRank or $k$-shell. Finally, since our algorithm is based on only local information, we claim that it can be efficiently applied to many large real systems.

\section{Empirical analysis and method}
\begin{figure}
\onefigure[width=8cm]{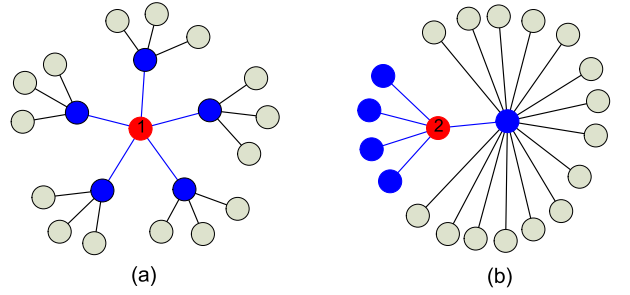}
\caption{The example of two toy networks. Red nodes in (a) and (b) have the same degree and the same number of second neighbors while the distributions of their neighbors' degree are different.}
\label{fig.1}
\end{figure}

To begin our analysis, we first discuss the diversity of the spreading paths. Actually, it is very difficult to trace all the spreading paths of each node. We therefore limit ourselves to study the diversity the local paths (i.e. the paths with length $2$). For better illustration, we give an example in fig.~\ref{fig.1}. The red nodes have the same degree and the same number of second nearest neighbors. The number of path for the target red node to infect each node is exactly the same in those two networks. However, all the paths to infect the gray nodes in fig.~\ref{fig.1}(b) overlap in the first half. The information can spread to gray nodes only if one specific blue node is infected. On the other hand, the paths to the gray nodes in fig.~\ref{fig.1}(a) is very diverse. The information may spread to gray nodes as long as any blue node infected. Intuitively, information could spread to gray nodes from the red node more easily in fig.~\ref{fig.1}(a) than in fig.~\ref{fig.1}(b).

In order to compute the probability of the gray nodes getting infected, we consider the SIR model \cite{Anderson-BOOK-1992} with infection rate $\mu$ in this toy network. It has been pointed out that in real case an individual cannot contact with all his/her neighbors \cite{Zhou-PRE-2006,Yang-PLA-2007,Gomez-EPL-2010}. Therefore, we assume that each node $i$ contacts $\sqrt{k_i}$ neighbors in each step (i.e. only $\sqrt{k_i}$ neighbors have the possibility $\mu$ to be infected by node $i$ in each step). In fig.~\ref{fig.1}, if we set the red nodes as the initial infected nodes, the expected number of infected nodes in fig.~\ref{fig.1}(a) and fig.~\ref{fig.1}(b) at the end will be $1+\sqrt{5}\mu+\sqrt{15}\mu^2$ and $1+\sqrt{5}\mu+\sqrt{3}\mu^2$, respectively. Apparently, the red node in fig.~\ref{fig.1}(a) can infect more nodes than that in fig.~\ref{fig.1}(b) simply due to the diverse paths.

In fact, the diversity of local paths can be represented by the degree evenness of the target nodes' neighbors. One can easily see that the degree of the red node's neighbors are more uneven in fig.~\ref{fig.1}(b) than that in fig.~\ref{fig.1}(a). To characterize such unevenness, we employ the information entropy \cite{Shannon1948} $h_i$ of each node $i$ in a network as
\begin{equation}
\label{eq.1}
h_i=\sum_{j\in\Gamma_i}-p_j log(p_j),
\end{equation}
where $p_j=\frac{k_j}{\sum_{l\in\Gamma_i}k_l}$, $\Gamma_i$ is the set of neighbors of node $i$, and $k_j$ is the degree of node $j$. A high value of $h_i$ indicates that the degree of neighbors of the target node is even, which corresponds to diverse paths for the target node to propagate information. However, there is a shortcoming in eq.~\ref{eq.1}: the entropy of two nodes with different degree will not be the same, even though the degrees of their neighbors are entirely even. Therefore, the entropy defined in eq.~\ref{eq.1} should be normalized according to the target node's degree so as to overcome this drawback,
\begin{equation}\label{eq.2}
    H_i=\frac{\sum_{j\in\Gamma_i}-p_jlog(p_j)}{\sum_{j\in\Gamma_i}-\frac{1}{k_i}log(\frac{1}{k_i})}=\frac{\sum_{j\in\Gamma_i}-p_jlog(p_j)}{log(k_i)}.
\end{equation}
In this way, the value of $H_i$ will be between $0$ and $1$, independent of target nodes' degree.

For directed network, we focus on the out-degree of each node. The eq.~\ref{eq.2} can be easily modified as,

\begin{equation}\label{eq.21}
    H_i^{out}=\frac{\sum_{j\in\Gamma_i^{out}}-p_jlog(p_j)}{log(k_i^{out})},
\end{equation}
where $\Gamma_i^{out}$ is the set of $i$'s followers who will receive information from $i$, $k_i^{out}$ is the out-degree of node $i$ (i.e., the number of followers of $i$) and  $p_j=\frac{k_j^{out}}{\sum_{l\in\Gamma_i^{out}}k_l^{out}}$.

Four real social networks including Youtube \cite{youtube-2012}, Orkut \cite{youtube-2012}, EmailEU \cite{email-2007} and Digg \cite{digg-2010} are used to empirically investigate the information entropy distribution of nodes. In these four real networks, Youtube and Orkut are undirected networks, EmailEU and Digg are directed ones. Youtube is a video-sharing website that includes a social network, in which users form friendship with each other and users can create groups which other users can join. Orkut is a free online social network where users form friendship with each other. Like youtube, Orkut also allows users form groups which other members can then join. EmailEU is generated by using email data from a large European research institution from October 2003 to May 2005. Given a set of email messages, each node corresponds to an email address and a directed edge between nodes $i$ and $j$ exists if $i$ sent at least one email message to $j$. Digg contains data about stories promoted to Digg's front page over a period of a month in 2009. The authors retrieved the voters' friendship networks where a node is corresponding to a user and a link $i\rightarrow j$ means node $j$ is watching the activities of (is a fan of) $i$. The number of nodes $N$, the largest degree $k_{max}$ (or the largest out-degree in the case of directed networks), the average degree $\langle k\rangle$ (or the average out-degree for directed networks), and the average clustering coefficient $\langle c\rangle$ are listed in table~\ref{tab.1}.

\begin{table}
\caption{Basic information on four real networks.}
\label{tab.1}
\begin{center}
\begin{tabular}{l rrrrr}\hline
Network  & $N$ & $k_{max}$ & $\langle k\rangle$ & $\langle c\rangle$\\ \hline
Youtube & 1,134,890 &  28,754 & 5.2650 & 0.1723\\
Orkut & 3,072,441 &  33,313 & 76.2814 & 0.1698\\
EmailEU & 265,214 &  929 & 1.5838 & 0.3093\\
Digg & 279,630 & 12,097  & 9.3623 & 0.0775\\ \hline
\end{tabular}
\end{center}
\end{table}

\begin{figure}
\onefigure[width=8cm]{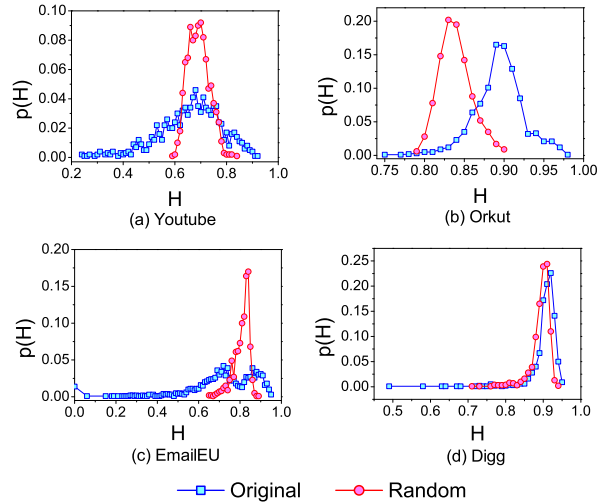}
\caption{The entropy distribution in four social networks and their corresponding random ones.}
\label{fig.2}
\end{figure}

For each network, we get the counterpart random networks according to the link swap method \cite{science.2002.maslov}. At each step, we randomly select a pair of edges $A-B$ and $C-D$. These two edges are then rewired to be $A-D$ and $B-C$. To prevent multiple edges connecting the same pair of nodes, if $A-D$ or $B-C$ already exists in the network, this random edge selection is aborted and a new pair of edges is randomly selected. We compare the distribution of $H$ in original network and the counterpart random networks, as shown in fig.~\ref{fig.2}. For simplicity, we only take into account the $1000$ highest degree nodes since they are more likely to have strong spreading ability than those with smaller degree. In fig.~\ref{fig.2}, one can see that the information entropy distribution of nodes in the original networks is much wider than that in the randomized ones, especially in Youtube and EmailEU networks. For some nodes in real network, their neighbors' degree can be very uneven. On the other hand, the neighbors' degree for some other nodes can be quite even. Therefore, the degree unevenness of the neighboring nodes shouldn't be neglected when ranking the spreaders.

Many researches already show that the degree centrality is not enough to accurately identify influential nodes \cite{Kitsak-ranking-2010,Chen-ranking-2011}. Actually, the ranking methods considering also the average degree of the neighboring nodes can effectively improve the pure degree method in ranking spreaders \cite{Chen-ranking-2011}. After taking into account the neighboring nodes' average degree, the score for each target node can be
\begin{equation}\label{eq.3}
    f_i=k_i+\lambda\frac{\sum_{j\in\Gamma_i}k_j}{k_i},
\end{equation}
where $\lambda$ is a tunable parameter.

We employ the Susceptible-Infected-Recovered (SIR) model \cite{Anderson-BOOK-1992} to simulate the spreading process on networks. This model is usually used to mimic the spreading processes where infected nodes will either get immunity and not infected again or die. Individuals in SIR model are classified in three classes according to their states: susceptible, infected and recovered. The simulation runs in discrete time steps. In a social network, a user neither contact all of her neighbors nor only a single neighbor, but part of her neighbors \cite{Zhou-PRE-2006,Yang-PLA-2007,Gomez-EPL-2010}. Moreover, the node with larger degree may contact more neighbors. Therefore, at each time step in our simulation, every infected node $i$ will select each of her neighbors (or followers) with probability $p$ ($p=\frac{1}{\sqrt{k_i}}$ in this letter) and then transmit the information to her with probability $\mu$ if the selected neighbor (or follower) is a susceptible one. The recovery rate is set as $1$ here. The simulation stops when there is no infected node anymore. Notice that this model is slightly different from the standard SIR model where all the followers of an infected node have the chance to be infected. The present mechanism is usually used to mimic the limited neighbor contact capability that is positive correlation to the degree of individuals \cite{Zhou-PRE-2006,Yang-PLA-2007}.

\begin{figure}
\onefigure[width=8cm]{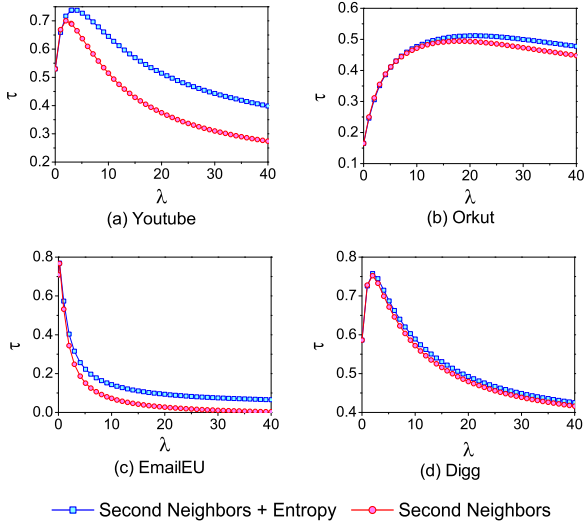}
\caption{Kendall's tau correlation coefficient $\tau$ between real spreading ability $s_i$ and the ranking score $f_i$ of the $1000$ largest degree nodes on four real networks when the simple ranking methods in eq.~\ref{eq.3} and eq.~\ref{eq.4} are applied. The results are averaged over 100 independent simulations.}
\label{fig.3}
\end{figure}

The number of nodes that are finally infected when the infection starts from a given node $i$ is denoted as its spreading ability $s_i^{\mu}$ where $\mu$ is the infection rate in the SIR model. Here, we select relatively small value of $\mu$ so that the infected percentage of the nodes is not so large and the network topology significantly affects the spreading result of nodes. Specifically, $\mu$ is set around the phase transition point of $\langle s^{\mu}\rangle$ in this letter, namely $\mu=0.04$ in Orkut network and $\mu=0.05$ in the other three networks. As mentioned above, we select $1000$ highest degree nodes from each network and calculate the Kendall's tau correlation coefficient ($\tau$) between $f_i$ and $s_i^{\mu}$. The results are reported in fig.~\ref{fig.3}. Clearly, $\tau$ increases with $\lambda$ at first, and then decreases after reaching a maximum. This indicates that the average degree of the neighboring nodes are beneficial for improving the identification of the most influential spreaders. However, when $\lambda$ gets very large, the node with small degree but with large average degree of neighbors will have high ranking. This would generally decrease the $\tau$ value.

We further move to improve the ranking accuracy by the path diversity. Specifically, we will make use of the entropy $H$ to modify eq. ~\ref{eq.3} as
\begin{equation}\label{eq.4}
    f_i^*=k_i+H_i\lambda\frac{\sum_{j\in\Gamma_i}k_j}{k_i}.
\end{equation}

The Kendall's tau correlation coefficient between $s_i^{\mu}$ and $f_i^{*}$ is also shown in fig.~\ref{fig.3}. One can see that the optimal $\tau$ can be even higher after $H_i$ is introduced.

So far, we have shown that the local path diversity is very valuable information for identifying influential nodes. For the local path number, the nodes' degree and the average degree of neighboring nodes are two key factors. Therefore, we combine these two information (local path diversity and number) and propose a local spreader ranking method named KED in this Letter. It can be described as
\begin{equation}\label{eq.5}
    f_i=k_iE_iD_i,
\end{equation}
where $E_i=1+H_i$ with $H_i$ defined in eq.~\ref{eq.2} for undirected network or $E_i=1+H_i^{out}$ with $H_i^{out}$ defined in eq.~\ref{eq.21} for directed network, and $D_i$ is a function of the average degree of the neighbors of node $i$. In this Letter, $D_i=e^{\frac{\sum_{j\in\Gamma_i}k_j}{N}}$. The range of value of $E_i$ and $D_i$ are $1\leq E_i\leq 2$ and $1\leq D_i\leq e$, respectively. In this ranking metric, the node with larger $f_i$ is supposed to have higher spreading ability. As an example, the $f_i$ for the red node is $25.9187$ in fig.~\ref{fig.1}(a) and $19.2212$ in fig.~\ref{fig.1}(b), which indicates that the red node in fig.~\ref{fig.1}(a) has stronger spreading ability than that in fig.~\ref{fig.1}(b). It is also noted that the model described in eq.~\ref{eq.5} is not only used to ranking nodes in undirected networks, but also in directed networks.

For comparison, we briefly describe two ranking algorithms on directed networks: PageRank \cite{Page-PR-1999} and LeaderRank \cite{Lv-leader-2011}, and one algorithm on undirected networks: $k$-shell \cite{Kitsak-ranking-2010}.  PageRank is depicted as a random walk on hyperlinked networks. The score $r_i(t)$ for node $i$ at time step $t$ is given by
\begin{equation}\label{EqPR}
    r_i(t)=\gamma+(1-\gamma)\sum_{j=1}^N\left[\frac{a_{ij}}{k_j^{in}}\left(1-\delta_{k_j^{in},0}\right)+\frac{1}{N}\delta_{k_j^{in},0}\right]r_j(t-1),
\end{equation}
where $k_j^{in}$ is the in-degree of node $j$ (i.e., the number of leaders of node $j$), parameter $\gamma$ is the probability for which a web page surfers to jump to a random web page, and for probability $1-\gamma$ a web page surfers to continue browsing through hyperlinks ($\gamma=0.15$ in our simulation), and $\delta_{k_j^{in},0}=1$ if $k_j^{in}=0$, otherwise $\delta_{k_j^{in},0}=0$.

LeaderRank is also a random-walk-based ranking algorithm \cite{Lv-leader-2011}. Different from PageRank, LeaderRank introduces a ground node $g$, which has two directed links $e_{gi}$ and $e_{ig}$ to every node $i$ in the original network. The score $r_i(t)$ of node $i$ at time $t$ is given by
\begin{equation}\label{EqFR}
    r_i(t)=\sum_{j=1}^{N+1}\frac{a_{ij}}{k_j^{in}}r_j(t-1).
\end{equation}

At the steady state, the score of the ground node is equally distributed to all other nodes to conserve scores on the nodes of interest. Therefore, the final score of node $i$ is defined as
\begin{equation}\label{EqFRLeadership}
   R_i=r_i(t_\infty)+\frac{r_g(t_\infty)}{N},
\end{equation}
where $r_i(t_\infty)$ is the score of node $i$ at the steady state according to eq.~\ref{EqFR}.

The $k$-shell decomposition is a deterministic ranking algorithm. The $k$-shell method starts by removing all nodes with one connection only (with their links), until no
more of such nodes remain, and assign them to the $1$-shell. After assigning the 1-shell, all nodes with residual degree
$2$ are recursively removed and the $2$-shell is created. This procedure continues as the residual degree increases until all nodes
in the networks have been assigned to one of the shells. For the details, readers could refer to \cite{kshell-Carmi-2007, Kitsak-ranking-2010}. In this Letter, if two nodes have the same $k$-shell value but different degrees, the node with larger degree will be set to a higher rank than the other one.

\section{Experiments and results}\label{exp}
The performance of the KED method will be tested in four real social networks and it will be compared with the degree centrality, $k$-shell decomposition, PageRank, and LeaderRank methods.

\begin{figure}
\onefigure[width=8cm]{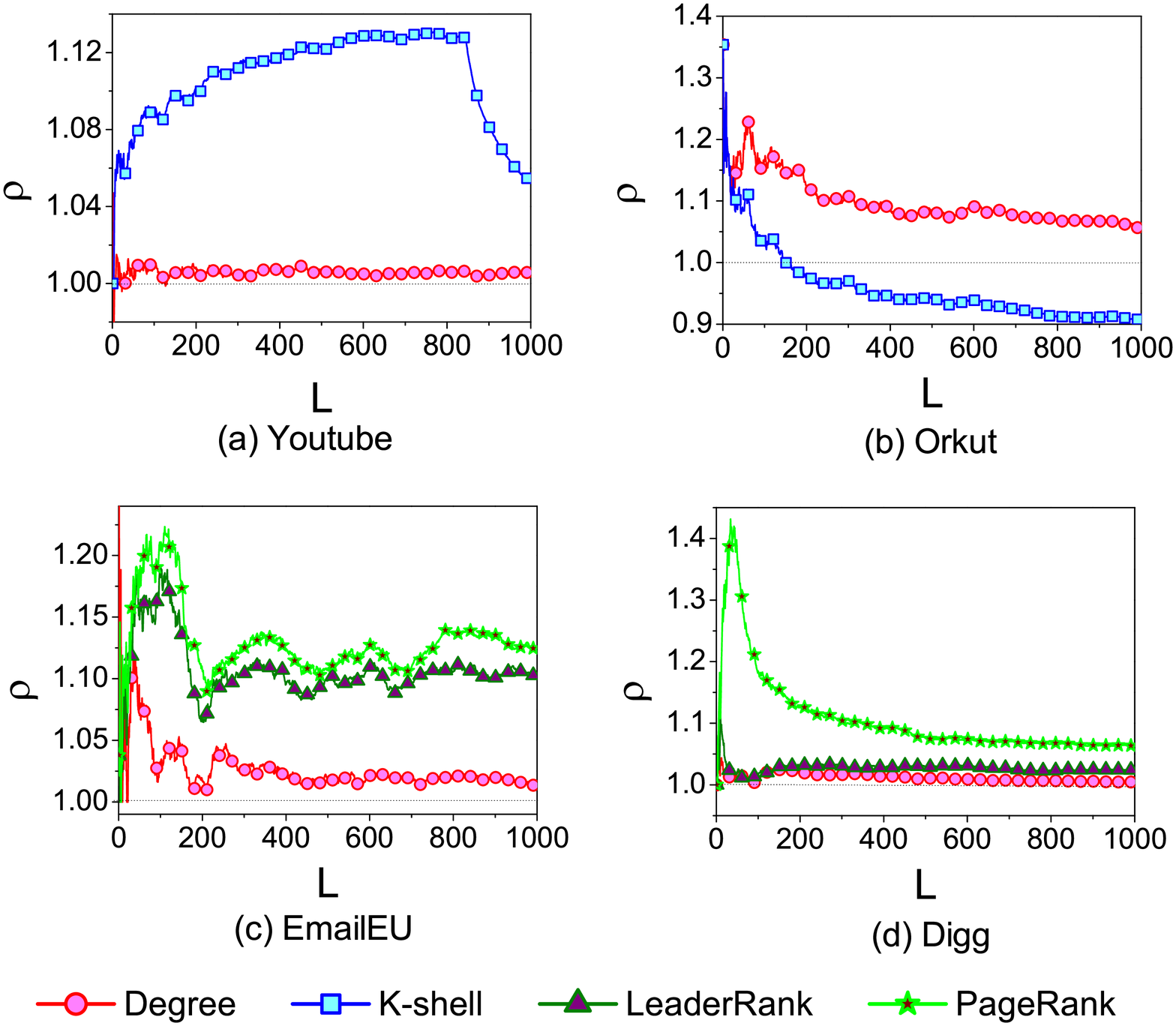}
\caption{The ratio $\rho$ of average final coverage $\langle s^{\mu}\rangle$ of top-$L$ nodes obtained by KED to that by degree, $k$-shell, PageRank or LeaderRank on four real networks. The spreading starts from each single node separately in the top-$L$ list. The results are averaged over 100 independent simulations.}
\label{fig.4}
\end{figure}

To investigate the influence of a node in information spreading, we initially set this node to be infected. As discussed above, the final coverage of this node $s_i^{\mu}$ is used to represent the spreading ability of $i$. The average final coverage $\langle s^{\mu} \rangle$ of top-$L$ ranked nodes obtained by each ranking algorithm is used to investigate the performance of these algorithms. We define $\rho$ as a ratio of $\langle s^{\mu} \rangle$ of KED to $\langle s^{\mu} \rangle$ of other methods. The results in fig.~\ref{fig.4} show that $\rho$ is larger than $1$ in four networks. It indicates that the information can spread wider from the top ranked nodes obtained by KED than that by degree, $k$-shell, PageRank and LeaderRank.

\begin{figure}
\onefigure[width=8cm]{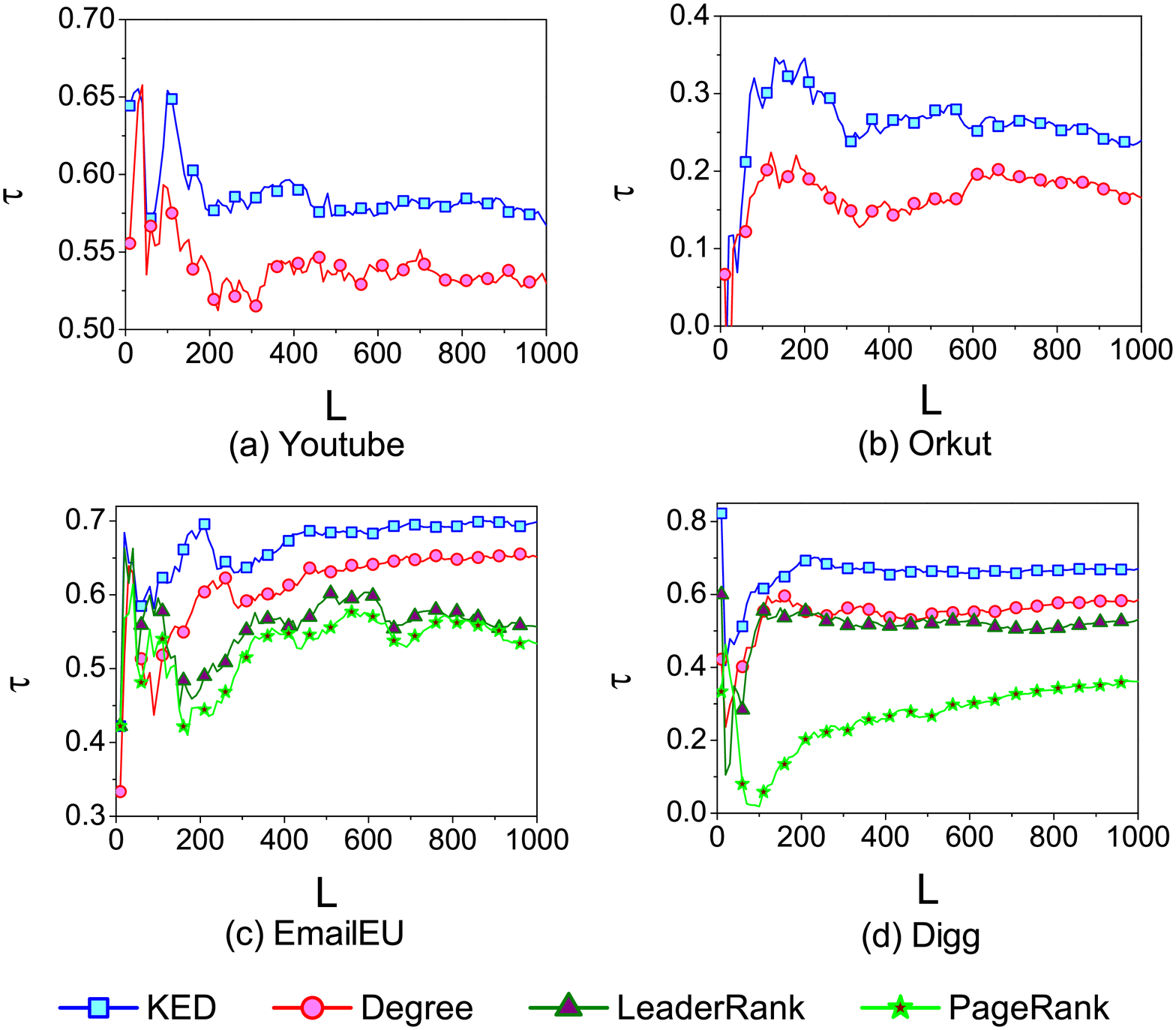}
\caption{Kendall's tau correlation coefficient $\tau$ between real spreading ability $s_i$ and the ranking score $f_i$ of the $1000$ largest degree nodes on four real networks. The results are averaged over 100 independent simulations.}
\label{fig.5}
\end{figure}

Kendall's tau correlation coefficient between the ranking of nodes and the real spreading ability is also calculated. We don't compare KED with $k$-shell since many nodes are with the same $k$-shell value, which makes $\tau$ value of $k$-shell method very low. Like before, we take into account $1000$ highest degree nodes. The Kendall's tau correlation coefficient between the ranking of top-$L (L\leq 1000)$ nodes by degree centrality, PageRank,  LeaderRank and KED and their real spreading ability is shown in fig.~\ref{fig.5}. One can see that generally $\tau$ of KED is the largest among all the ranking methods. That is to say, KED ranking algorithm can generate a more accurate ranking of spreading ability than degree centrality, PageRank and LeaderRank.

\begin{figure}
\onefigure[width=8cm]{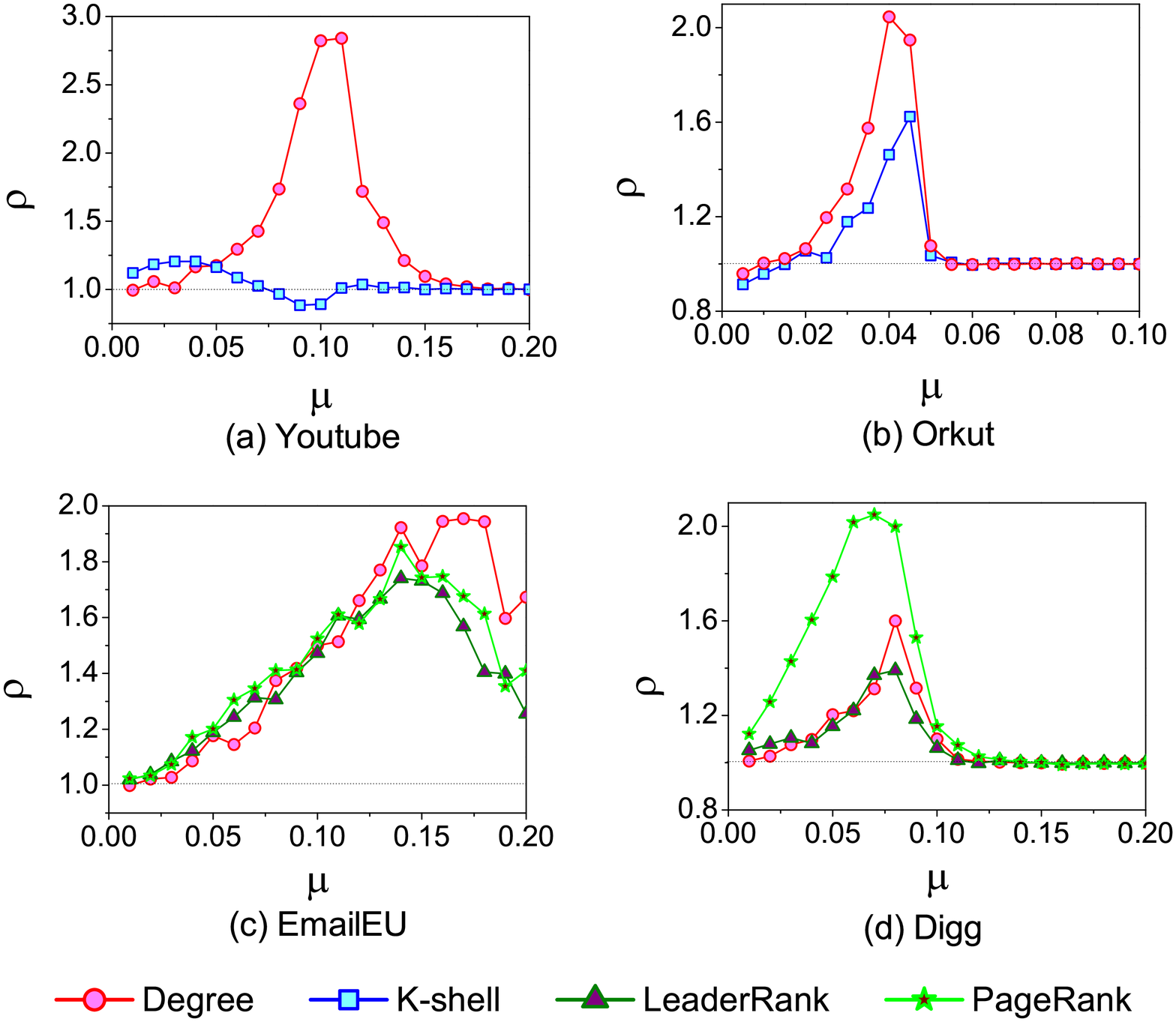}
\caption{The ratio $\rho$ of average final coverage $\langle s^{\mu}\rangle$ of top-$50$ nodes obtained by KED to that by degree, $k$-shell, PageRank or LeaderRank on four real networks. All the $50$ nodes are set as the initial spreaders in each simulation. The results are averaged over 100 independent simulations.}
\label{fig.6}
\end{figure}

Figs.~\ref{fig.4} and \ref{fig.5} indicate that the KED ranking algorithm performs well in ranking the spreading ability of individual nodes. It is already pointed out that a good ranking method for individual nodes might not be effective in identifying the group of nodes with strongest spreading ability \cite{Kitsak-ranking-2010}. This is because the influential nodes identified by the method might densely connected to each other, which makes the virus/information propagate in a small region. Therefore, the performance of different algorithms on identifying the group of nodes with strong spreading ability is also investigated. We choose the top-$K$ ($K = 50$ here) most influential nodes from each algorithm. Since there might be a considerable number of overlapped nodes in top-ranked lists of two algorithms, we compare the spreading processes resulted from those non-overlapped nodes in the top-ranked lists. That is, each time when we compare the KED and another algorithm, the nodes appeared in only one list are set to be the initially infected ones. Fig.~\ref{fig.6} shows the ratio $\rho$ of the number of final infected nodes resulted from top-ranked nodes by KED to those by other ranking algorithms at different infected rates $\mu$. In fig.~\ref{fig.6}, a $\rho$ higher than $1$ indicates that KED method outperforms another method. Clearly, it shows in fig.~\ref{fig.6} that the performance of KED is almost better than all other ranking algorithms. Note that in the case of large $\mu$ values, where spreading can cover almost all the network, the role
of initial infected nodes is no longer important and the final coverage of virus is independent of where it originated from. Therefore, $\rho$ in fig.~\ref{fig.6} converges to $1$ when $\mu$ is big.

\section{Conclusion}
In this Letter, we introduced the concept of \emph{path diversity} in ranking spreaders. Measured by information entropy, it is introduced to a simple spreader ranking algorithm. The results show that not only the number of paths that determines the spreading ability of a node, the path diversity is also a significant factor. We further proposed a local spreader ranking algorithm named KED, in which the information of path number and path diversity are combined. We use SIR model to simulate the spreading process on four real social networks. The results show that the performance of KED is much better than that of degree centrality, PageRank and LeaderRank, and in most cases, is better than that of $k$-shell. We show that KED is not only good at identifying the single influential spreader, but also very effective in finding the group of nodes with strong spreading ability. Moreover, KED is an universal method which is suitable for both directed and undirected networks. From the practical point of view, it can be easily applied to very large real systems since it is based on only local information. Actually, how to identify influential nodes in networks, especial in temporal networks and bipartite networks, is a long term challenge problem. Some progress has been made in this direction\cite{Ghosh-temporal-2011,ZhouYB-2012,WeiDJ-2013,ZhangXH-2013}, but systematic analysis is still lacking. We remark here that in these systems, the combination of path number and path diversity may also lead to an improvement in ranking spreaders.

\acknowledgments
This work was partially supported by the National Natural Science Foundation of China under Grant Nos. 61103109 and 61003231.  D.B.C. acknowledges the Huawei university-enterprise cooperation project YBCB2011057.

\end{document}